\documentclass[conference]{IEEEtran}
\RequirePackage{etoolbox} 
\IEEEoverridecommandlockouts

\usepackage{cite}
\usepackage{amsmath,amssymb,amsfonts}
\usepackage{algorithmic}
\usepackage{graphicx}
\usepackage{textcomp}
\usepackage{xcolor}
\def\BibTeX{{\rm B\kern-.05em{\sc i\kern-.025em b}\kern-.08em
    T\kern-.1667em\lower.7ex\hbox{E}\kern-.125emX}}
    \RequirePackage{etoolbox} 

\usepackage{algorithm}
\usepackage{algorithmic}

\usepackage{subcaption}
\usepackage{xcolor,soul}
\usepackage{graphicx}
\usepackage{tikz}
\usepackage{tikz-qtree}
\usepackage{graphics}
\usepackage{pgfplots}
\pgfplotsset{compat=1.18} 
\usepgfplotslibrary{groupplots,fillbetween}
\usetikzlibrary{backgrounds,calc,shapes,arrows,arrows.meta,shapes.arrows,shapes.symbols,patterns, datavisualization, datavisualization.formats.functions,matrix,decorations.markings,positioning,shapes,fit,arrows,intersections, decorations.text}
\usepackage{dsfont}
\usepackage{amsthm}
\usepackage{amsfonts}
\usepackage{mathtools}
\usepackage[short]{optidef}
\newtheorem{proposition}{Proposition}
\usepackage{float}
\floatstyle{plain}
\newfloat{floatequation}{htbp}{loe}
\floatname{floatequation}{Equation}

\AfterEndPreamble{
  \mathchardef\mathcomma\mathcode`\,
  \mathcode`\,="8000 
}
{\catcode`,=\active
  \gdef,{\mathcomma\discretionary{}{}{}}
}
\begin{document}

\title{Route Planning and Online Routing for Quantum Key Distribution Networks\\
\thanks{This work was partially funded by the European Commission project number 101091675, FranceQCI through its DIGITAL Action Grant.}
}

\author{\IEEEauthorblockN{Jorge L\'opez, Charalampos Chatzinakis, and Marc Cartigny}
\IEEEauthorblockA{\textit{Airbus}\\
Boulogne-Billancourt, France \\
\{jorge.lopez-c,charalampos.chatzinakis, marc.cartigny\}@airbus.com}
}

\maketitle

\begin{abstract}
Quantum Key Distribution~(QKD) networks harness the principles of quantum physics in order to securely transmit cryptographic key material, providing physical guarantees. These networks require traditional management and operational components, such as routing information through the network elements. However, due to the limitations on capacity and the particularities of information handling in these networks, traditional shortest paths algorithms for routing perform poorly on both route planning and online routing, which is counterintuitive. Moreover, due to the scarce resources in such networks, often the expressed demand cannot be met by any assignment of routes. To address both the route planning problem and the need for \emph{fair} automated suggestions in infeasible cases, we propose to model this problem as a Quadratic Programming~(QP) problem. For the online routing problem, we showcase that the shortest (available) paths routing strategy performs poorly in the online setting. Furthermore, we prove that the widest shortest path routing strategy has a competitive ratio greater or equal than $\frac{1}{2}$, efficiently addressing both routing modes in QKD networks.

\end{abstract}


\input{images/TIKZ_network_drawing}

\section{Introduction}
Currently, secure communications rely on different encryption algorithms whose security guarantees rely on the computational hardness of finding a secret key. Such elegant computing-based security schemes are not a 100\% safe. The main problem is that even if finding a key by brute-force requires an exponential amount of guesses (w.r.t. the length of the secret key), this is only in the worst case; this implies that a lucky attacker may find the key in less time. Furthermore, birthday algorithms, frequency analysis, and many other side-channel attacks are some common means to compromise state-of-the-art secure communications.

Quantum communications offer an interesting promise from the security standpoint. Due to the fact that observing a quantum channel produce noticeable changes in the data transmitted \cite{bb84++}, and thus, an attacker cannot intercept any communications without it being detected. Furthermore, due to the no-cloning theorem, the attacker cannot re-inject a copy of the read information. This implies that security guarantees are given by quantum mechanics, which cannot be altered. Quantum Key Distribution~(QKD) relies on quantum communications, and it aims at transmitting secure information though a network of nodes with quantum channels. A typical centralized QKD architecture \cite{qkdsdn} is depicted in Figure~\ref{fig:qkdnet}; industrial actors have adopted such design \cite{oqtavo}. In this figure, quantum links are depicted with solid black lines while regular communication links are depicted in red dashed lines. Data is transmitted across the network by sending a ciphertext xor (denoted $\oplus$) based on the product of two keys exchanged through a quantum channel. Figure~\ref{fig:qkdnet} also illustrates this transmission. Note that, if no keys can be intercepted in the quantum channels, retrieving the original data ($k_A$) can only be done if both a transit node and the centralized quantum key manager are compromised. 

\begin{figure}[!htb]
\centering
    \begin{tikzpicture}[scale=.75, every node/.style={scale=.75}]
    \node[server]                               (h1)    {};
    \node               at ([yshift=.7cm]h1)    (h1l)   {$A$};
    \node[l3 switch]    at ([xshift=2cm]h1)     (s1)    {};
    \node               at ([yshift=-0.5cm]s1)     (s1l)   {$q_1$};
    \node[l3 switch]    at ([yshift=1cm,xshift=2cm]s1)     (s2)    {};
    \node               at ([yshift=-0.5cm]s2)     (s2l)   {$q_2$};
    \node[l3 switch]    at ([yshift=-1cm,xshift=2cm]s1)  (s3)    {};
    \node               at ([yshift=-0.5cm]s3)    (s3l)   {$q_3$};
    \node[l3 switch]    at ([xshift=4cm]s1)     (s4)    {};
    \node               at ([yshift=-0.5cm]s4)    (s4l)   {$q_4$};
    \node[server]       at ([xshift=2cm]s4)     (h4)    {};
    \node               at ([yshift=.7cm]h4)    (h4l)   {$B$};
    \node[rectangle,draw]       at ([yshift=1.5cm]s2)     (QKM)   {\textcolor{white}{---}Quantum Key Manager (QKM)};
    
    \draw[thick,red, dashed] (h1)--(s1);
    \draw[thick,red, dashed] (h4)--(s4);
        
    \draw[thick] (s1.east)--(s2.west);
    \draw[thick] (s1.east)--(s3.west);
    \draw[thick] (s2.east)--(s4.west);
    \draw[thick] (s3.east)--(s4.west);
    
    \draw[very thick, red, dashed]    ([yshift=0cm,xshift=0cm]s1.north)-- node [left] {}  ([yshift=0cm,xshift=0.25cm]QKM.south west);
    
    \draw[very thick, red, dashed]    ([yshift=0cm,xshift=-0]s2.north)-- node [left] {}  ([yshift=0cm,xshift=-0cm]QKM);
    
    \draw[very thick, red, dashed]    ([yshift=0cm,xshift=0.5cm]s3.north)-- node [right] {}  ([yshift=0cm,xshift=0.5cm]QKM);
    
    \draw[very thick, red, dashed]    ([yshift=0cm,xshift=-0.25cm]QKM.south east)-- node [right] {}  ([yshift=0cm]s4.north);
    
    \draw[-Latex,very thick, gray, dashed]    ([yshift=0.2cm,xshift=0cm]h1.east)-- node [above] {$k_A$}  ([yshift=0.2cm]s1.west);
    
     \draw[-Latex, very thick, gray, dashed]    ([yshift=0cm,xshift=-0.2cm]s1.north)-- node [left] {$k_A\oplus k_{1,2,i}$}  ([yshift=0cm,xshift=-0cm]QKM.south west);
    
    \draw[-Latex,very thick, gray, dashed]    ([yshift=0cm,xshift=-0.2cm]s2.north)-- node [left] {$k_{1,2,i}\oplus k_{2,4,j}$}  ([yshift=0cm,xshift=-1cm]QKM);
    
    \draw[-Latex,very thick, gray, dashed]    ([yshift=0cm,xshift=0cm]QKM.south east)-- node [right] {$k_A\oplus k_{1,2,i}\oplus k_{1,2,i}\oplus k_{2,4,j}$}  ([yshift=0cm,xshift=0.2cm]s4.north);
    
    \draw[-Latex,very thick, gray, dashed, shorten <= 0.2cm, shorten >= 0.5cm]    ([yshift=0.2cm,xshift=0cm]s4.east)-- node [above,pos=0.3] {$k_A$}  ([yshift=0.2cm]h4.west);
\end{tikzpicture}
    \caption{Example QKD network and transmission}
    \label{fig:qkdnet}
\end{figure}

However, how is the data path chosen? In the previous example, why not use node $q_3$ as relay? This is our problem of interest, and is related to the control plane, where routing decisions are made in such QKD networks. From the routing perspective, interesting challenges arise. Traditional shortest-paths algorithms used to route traffic in general communication networks (including the well-known Open Shortest Path First  -- OSPF protocol) can yield poor solutions for QDK networks as certain short paths may be overloaded. We thus explore two different routing modes in QKD networks. First, we assume that there exists a predefined demand for each pair of nodes, and therefore, the task is a route planing task. However, due to the limitations in quantum equipment \cite{qlinkperf}, it is interesting to consider automated suggestions for unfeasible instances, i.e., what to do when we cannot meet the demand. The second possibility is to have spontaneous demands of key exchanges for a given pair of source and destination nodes.

In the literature, different algorithms have been proposed to route quantum keys in QKD networks. However, most of them are based on shortest paths \cite{qkdssrouting, qkdssrouting2, qkdssrouting3}. Others proposed advanced schemes based on demand \cite{qkdroutedemand} or on current resources and the probability of future arrivals \cite{qkdprobarouting}. However, all the previous works evaluate the performance of the proposed approach based on simulations, providing no mathematical guarantees on the performance of the proposed approaches;  providing such guarantees is the goal of this work.

\section{Modeling of the problems of interest}
Our problems of interest are based on transport networks, and thus, we model them evidently as a directed and weighted graph $N=(V,E,k)$, where: $V$ is a set of nodes; $E\subseteq V\times V$ is a set of edges (ordered pairs of nodes); $k: E\to \mathds{Z}_{+}$ represents a key material refresh rate capacity (for easiness we refer to it as bandwidth) for the route planning problem or a buffer size capacity for on-demand routing. A path $\pi$ is a sequence of edges ($\pi\in E^*$); the empty path is denoted as $\epsilon$). We denote $\pi_i$ the $i$-th edge of the path and $|\pi|$ its length. $\pi$ is simple if all nodes in the path are distinct. $\pi$ is valid if it is connected, i.e., $\forall i\in\{1,\ldots,|\pi|-1\}\;dst(\pi_i)=src(\pi_{i+1})$. A very simple example is the graph shown in Figure~\ref{fig:example_topology}, where the nodes are $V=\{Q_1,Q_2,Q_3\}$, the edges (links) are $E=\{(Q_1,Q_2),(Q_1,Q_3),(Q_2,Q_1),(Q_2,Q_3),(Q_3,Q_1),(Q_3,Q_2)\}$, and the weights over the edges representing the \emph{resource} of interest (i.e., the number of keys in the buffer, e.g., for the link $(Q_3,Q_1)$ there are 3 bits present) are given by the function: $$k(i,j)=\begin{cases}1;\text{ if }(i,j)=(Q_2,Q_1)\\3;\text{ if }(i,j)=(Q_2,Q_3)\vee(i,j)=(Q_3,Q_1)\\2;\text{ otherwise}\end{cases}$$.

\begin{figure}[!htb]
    \centering
    \begin{tikzpicture}[node distance=2.5cm,>=stealth',bend angle=45,auto, scale=.8, every node/.style={scale=.8}]
    \tikzstyle{directed}=[->]
    \tikzstyle{node}=[circle,thick,draw=black!75, inner sep=0.075cm]
   
            \node[node]                 (q1)       {};
            \node[]                     (q1text)    at ([yshift=0.4cm]q1.center)    {$Q_1$};
            \node[node]                 (q2)        at ([xshift=2cm]q1)  {};
            \node[]                     (q2text)   at ([yshift=0.4cm]q2.center)  {$Q_2$};
            \node[node]                 (q3)        at ([yshift=-2cm]q1)  {};
            \node[]                     (q3text)   at ([yshift=-0.4cm]q3.center)  {$Q_3$};
            
            \path   
            (q1)    edge[directed, bend left=10]    node[above] {2}   (q2)
                    edge[directed, bend right=10]    node[above, rotate=90] {2}   (q3)
            (q2)    edge[directed, bend left=10]    node[below] {1}  (q1)
                    edge[directed, bend right=10]    node[above, rotate=45] {3}   (q3)
            (q3)    edge[directed, bend right=10]    node[above, rotate=-90] {3}   (q1)
                    edge[directed, bend right=10]    node[below, rotate=45] {2}   (q2)
            ;
\end{tikzpicture}
    \caption{Example topology}
    \label{fig:example_topology}
\end{figure}

The main goal of the problems of interest is to plan different path according to a form of demand / request. Therefore, we define the following objects of interest. For the route planning problem, we define a contract $c$ which is a four-tuple $c=(s,d,b,p)$, where $s,d\in V$, are source ($s$) and destination ($d$) nodes, $b\in\mathds{Z}_{+}$ is the requested bandwidth, and finally $p\in\mathds{Z}_{+}$ is a relative priority associated to the contract. The priority is a parameter that is useful for providing automated suggestions and ensuring that higher priority contracts get preferential treatment. Note that, for convenience if $c=(s',d',b',p')$, we denote $s(c)=s'$, $d(c)=d'$, $b(c)=b'$, and $p(c)=p'$. With respect to the online routing, requests are modeled similarly to contracts, however, there is no priority notion. Thus, for a request $r=(s,d,b)$. Additionally, there's a maximal number of bits that can be requested, a constant $\mu$. Consider our example topology (in Figure~\ref{fig:example_topology}), a set of contracts might be $C=\{(Q_1,Q_2,2,1),(Q_2,Q_1,3,3,10),(Q_2,Q_3,2,100)\}$ (which exceeds the available resources). With these objects we can then define our problems of interest.

We couple the route planning problem to the automated suggestions for unfeasible instances as route planning can be seen as a constraint satisfiability problem and providing automated suggestions can be seen as the optimization version of this problem, whenever not all contracts can be granted. We focus on a \emph{fair} version of automated suggestions, where fully honoring some contracts while disregarding others is considered to be undesirable or unfair. However higher priority contracts are more likely to be respected as the suggestions should weight the contract priority.

\paragraph{The Route Planning with Fair Automated Suggestions for Unfeasible Instances~(RPFASUI) Problem} Given a network $N=(V,E,k)$, and a finite set of QKD contracts $C$, also known as Service Level Agreement~(SLA), find a path for each contract (mapping solution) $\mathcal{S}: C\to E^*$, such that: i) the paths are simple and valid for the contracts and network; ii) the bandwidth of the allocated contracts does not exceed the bandwidth capacity of each link; ii) the difference between each of the contracts requested bandwidth and the granted bandwidth is minimal, fairly distributed across contracts and weighted by contract priority.

On the other hand, on-demand QKD routing occurs whenever either a key request is not planned in the SLA, or there does not exist an SLA for the routing of QKD key material. In this case, the algorithm responsible for path selection is unaware of any future demands. Furthermore, we assume that a (single) request must be served if there exists a feasible path that can transport this key, i.e., the algorithm cannot ``save'' quantum keys (stored at the nodes) for later as there is no knowledge of what is to come. Let us assume that a request is composed of a single key with a source $s\in V$, a destination $d\in V$, and the number $n\in\mathds{Z}_{+}$ of bits of the key. Finally, there is an important working assumption. Requests can have an arbitrary number of bits, however, in all practicality, the requested amount cannot be \emph{large}, or at least it is much smaller than the size of link buffers. Let $\mu$ denote this maximal number of requested bits and $\beta$ the size of the buffered bits in a quantum link; then $\beta\gg\mu$. With this in mind, we can define our problem of interest.

\paragraph{The Single Online Routing Request~(SORR) problem} Given a network $N=(V,E,k)$, a maximal key size request $\mu$, and QKD request $r=(s,d,n)$, where $s,d\in V$ and $n\leq \mu$, find a path for the request starting at $s$, finishing at $d$ and i) the path is simple and valid for the requests and network and ii) the key buffer in each of the links of the assigned path has at least $n$ bits i.e., find an assignment $r\mapsto \pi$, where $\pi\in E^*$; if no feasible path is found $\pi=\epsilon$.

Having both of our problems of interest define, in the next sections, we propose different methods to solve the aforementioned problems.

\section{Route planning with fair Automated suggestions for unfeasible instances}\label{sec:sla}
As previously discussed the RPFASUI problem is a combinatorial optimization problem. As such, we define the variables and objects of interest. We denote $c_i$ as the $i$-th contract in lexicographical order of $C$. Correspondingly, we denote $\mathbf{paths}(c_i)$ as a set of simple and valid paths for contract $c_i$; for convenience, we denote the $m$-th path in lexicographical order for the set of paths for contract $c_i$ as  $\mathbf{paths}(c_i)_m$. Let $\rho_{i,m}\in\{0,1\}$ be a binary variable denoting if the $i$-th contract should be routed via the $m$-th path. As potentially not all of the demand can be granted, we define the variables $\beta_i\in\mathds{Z}_{\geq0}$ representing the granted bandwidth for the contract $i$. Associated to the previous variables, we define $\beta_{i,m}\in\mathds{Z}_{\geq0}$ representing the granted bandwidth for the contract $i$ in path $m$. Likewise, let $\varepsilon_{i,j,l}\in\mathds{Z}_{\geq0}$ be a variable representing the granted bandwidth for the $i$-th contract on edge $(j,l)$. Finally, for convenience we define the characteristic function $\mathbf{1_{i}}((j,l))$ indicating if edge $(j,l)$ belongs to any of the paths in $\mathbf{paths}(c_i)$. Having defined the objects of interest, we can express the RPFASUI problem using a Quadratic Programming~(QP) formulation as shown in Equation~\ref{eq:gen_qp}. 

The constraints of the RPFASUI formulation are somewhat straightforward. The first family of constraints i) says that each of the contracts must at most one assigned path. This is due to the fact that, as explained before, keys are not transmitted through the network but, rather xor masks are sent to the centralized node. Having different paths would require the careful synchronization of partial key material which is impractical for the implementation of QKD protocols found in the state of the art. The second family of constraints ii) states that the granted bandwidth for the contract $i$ on path $m$ should be less or equal than the requested one, and should be 1 only if the path $m$ is the path assigned to contract $i$. The third family of constraints states that granted amount of bandwidth for contract $i$ on edge $(j,l)$ should equate to sum of all the granted bandwidth for the contract $i$ on all paths that cross that edge and they are assigned to contract $i$ (as there is a single path chosen this sum has a single term that can contribute to it). The fourth family of constraints iv) are simply capacity constraints. The final family of constraints v) states that the granted amount of bandwidth for flow $i$ must be equal to the sum of all the granted quantities for all possible paths (a single term contributes to this sum). We note that the problem can have an alternative formulation instead using paths using the flow conservation constraints from the  multi-commodity flow problem \cite{mcf}. However, we note that for practical implementations formulating problem with an explicit list of paths allows to consider a subset of pertinent paths, e.g., only paths of length at most three (and therefore yielding easier instances), although in the worst case its size may be exponential; this restrictions are dependant on operators, and other parameters as the quality of experience of end users. 

{\small
    \begin{align}
        &\text{\textbf{minimize} } \sum_{i=1}^{|C|}p(c_i)\cdot (b(c_i)-\beta_i)^2 \notag\\
        &\text{\textbf{subject to}: } \sum_{m=1}^{|\mathbf{paths}(c_i)|}\rho_{i,m} \leq 1;\; \forall i \in\{1,\ldots,|C|\} \text{ i)}\notag\\
        &\beta_{i,m} \leq b(i)\cdot \rho_{i,m};\; \forall i \in\{1,\ldots,|C|\} \forall m \in\{1,\ldots,\mathbf{paths}(c_i)\}\text{ ii)}\notag\\
        &\varepsilon_{i,j,l} = \sum_{m=1}^{|\mathbf{paths}(c_i)|} \mathbf{1_{i,m}}(j,l)\cdot\beta_{i,m};\; \forall i \in\{1,\ldots,|C|\} \forall (l,j)\in E\text{ iii)}\notag\\
        &\sum_{i}^{|C|}\varepsilon_{i,j,l} \leq k((j,l));\; \forall (j,l)\in E\text{ iv)}\notag\\
        &\beta_i = \sum_{m=1}^{|\mathbf{paths}(c_i)|}\beta_{i,m};\; \forall i \in\{1,\ldots,|C|\}\text{ v)}\label{eq:gen_qp}
    \end{align}
}

The objective function simply states that the difference between the demanded bandwidth and the granted one should be minimal for each of the contracts, and larger differences are discouraged (as the difference is squared) effectively distributing a fair granting scheme. However, larger priority contracts will only be affected whenever the equivalent amount in priority of lower priority contracts cannot be granted. We propose one formulation which is pertinent for the traditional industrial environment where contracts can have different priorities but we aim at still distributing fairly the resources within the same priority clusters. We note however, that different formulations can be possible as the objective function.

\subsection{Discussion: on different fair metrics (objectives)} 
There are different ways to define the concept of fairly distributing the resources of QKD networks among the contracts. One approach is to minimize the differences between the requested bandwidth and the allocated bandwidth. This method naturally penalizes larger discrepancies, ensuring that all contracts are fulfilled as much as possible by distributing any shortfall evenly. We call this metric the \emph{Evenly Spread Contract Fulfillment} (ESCF). Since the quantities involved are integers, a lack of resources will result in an increase of two capacity units only if at least four contracts have been fully allocated, three units if at least nine contracts are fully allocated, and so on. This behavior is modeled by the function $(b(c_i) - \beta_i)^2$ which penalizes larger differences more heavily. This implies that when resources are insufficient for all requested contracts, the shortfall will be distributed across multiple contracts, rather than disproportionately affecting just one. As an example, consider the topology shown in Fig.~\ref{fig:example_topology}, and the set of example contracts defined earlier: $C = {(Q_1, Q_2, 2, 1), (Q_2, Q_1, 3, 10), (Q_2, Q_3, 2, 100)}$. An optimal ESCF-based assignment would allocate two units to the second contract (in lexicographic order) and one unit to the third. However, a \emph{Prioritized Evenly Spread Contract Fulfillment} (PESCF), as defined in Equation~\ref{eq:gen_qp}, instead allocates one unit to the second contract and two units to the third. 

On the other hand, the notion of fairness might not be to equally distribute the shortfall across all contracts, but rather to ensure that the granted ratios (i.e., the proportion of each request that is fulfilled) are as similar as possible. As an example, consider one contract requesting 11 units and another requesting 31. Subtracting two units from each yields granted ratios of approximately 82\% and 94\%, respectively. However, subtracting one unit from the first and three units from the second results in granted ratios of approximately 91\% and 90\%. This latter distribution might be perceived as fairer, since the granted ratios are more closely aligned. Based on this intuition, we propose modeling the objective function by minimizing the squared differences between all pairs of granted ratios: $\sum_{i=1}^{|C|-1}\sum_{j=i+1}^{|C|}(\frac{\beta_i}{b(i)}-\frac{\beta_j}{b(j)})^2$. The goal is thus to maximize total allocated bandwidth while balancing the granted ratios across contracts. We note that due to constraint ii), the granted quantities are upper-bounded by their corresponding requests, ensuring consistency. The complete objective function becomes: $\sum_{i=1}^{|C|-1}\sum_{j=i+1}^{|C|}(\frac{\beta_i}{b(i)}-\frac{\beta_j}{b(j)})^2-\sum_{i=1}^{|C|}\beta_i$. We refer to this metric as the \emph{Evenly Distributed Granted Ratios} (EDGR).

We note that selecting an appropriate fairness metric for handling infeasible instances may depend on the specific application. However, based on practical experience, the two previously discussed schemes (ESCF and EDGR) represent the most commonly used notions of fairness.

\section{Strategies for the Online Routing Problem}\label{sec:online}
Note that solving the SORR problem can be achieved in different manners. For instance, choose the shortest path, share the load between all feasible paths, and many others. Furthermore, note that keys are being regenerated, and if the consumption rate is lower than the refresh rate then, keys will be refilled to nominal capacity (bounded by the buffer size). However, we do not consider these aspects as we are interested in the worst-case scenario, in order to provide worse-case guarantees. Let us first consider what is the worst-case performance of a shortest \emph{available} (i.e., shortest path where each node has keys stored in its buffer) paths algorithm for the SORR problem. 

\begin{proposition}
The shortest available path algorithm has a competitive ratio of $\frac{1}{1+\mu\lfloor\frac{|E|}{2}\rfloor}$.
\end{proposition}

\begin{proof}

(Direct proof.) When employing the shortest path algorithm in a QKD network instance $N=(V,E,k)$, in the worst case it will affect as many QKD requests as possible. In order for the shortest path to impact the maximum number of other QKD requests, it should take all available keys in a path where many other assignments could have been done by an optimal algorithm. This can be easily achieved if QKD requests exist per each link used by the shortest path assignment. 

Consider the QKD network instance depicted in Figure~\ref{fig:spworstcasegraph}, where for all links the key capacity is constant, i.e., $k(e)=\beta$. Note that the larger the key requests accepted by the shortest available path algorithm, the larger the number of failed request there are (as fewer request can be served, and thus, a larger fail ratio). In order to fail as much requests as possible, the idea is that choosing a single shortest path takes away resources from other requests, as many as possible. In this case, first, we should make the shortest path as long as possible, so that it affects as many requests as possible. This happens when the shortest path has a length of $\lfloor\frac{|E|}{2}\rfloor$ (a longer path will imply that this is not the shortest available.) This is evidently the case for networks with an odd number of edges, as the alternative path must have $\lfloor\frac{|E|}{2}\rfloor+1$, which is not the shortest one. 

Additionally, consider that for each edge in the shortest path there are $\beta$ requests per link after $\frac{\beta}{\mu}$ requests of $\mu$ bits from $s$ to $d$. That is the sequence: {\small $\overbrace{(s,d,\mu)\ldots(s,d,\mu)}^{\frac{\beta}{\mu} \text{ times}}\overbrace{\overbrace{(s,a,1)\ldots(s,a,1)}^{\beta \text{ times}}\ldots\overbrace{(a'd,1)\ldots(a',d,1)}^{\beta \text{ times}}}^{\beta\lfloor\frac{|E|}{2}\rfloor \text{ times}}$}. Note that this sequence would be completely served by an optimal algorithm, first it would use the longest path to serve the first $\frac{\beta}{\mu}$ requests, and then, each individual request out of the $\beta$ requests per link. For that reason, the competitive ratio of the shortest path algorithm is $\frac{\frac{\beta}{\mu}}{\frac{\beta}{\mu}+\beta\lfloor\frac{|E|}{2}\rfloor}$, simplifying the previous expression, we obtain that the shortest path algorithm has a competitive ratio of $\frac{1}{1+\mu\lfloor\frac{|E|}{2}\rfloor}$. \qedhere

\begin{figure}[!htb]
    \centering
    \begin{tikzpicture}[node distance=1cm,>=stealth',bend angle=45,auto]
            \tikzstyle{node}=[circle,thick,draw=black!75, inner sep=0.075cm]
            \tikzstyle{normalArrow}=[thick,->]
            \tikzstyle{dashedArrow}=[thick,->,dashed]
            \tikzstyle{dottedArrow}=[thick,->,dotted, gray!80!white]

            \node[node]                 (s)       {};
            \node[]                     (stext)    at ([yshift=0.3cm]s.center)    {$s$};
            \node[node]                 (n2)        at ([xshift=1cm,yshift=1cm]s)  {};
            \node[]                     (n2text)   at ([yshift=-0.3cm]n2.center)  {$a$};
            \node[]                     (topdots)  at ([xshift=1cm]n2)  {\textcolor{gray!80!white}{\ldots}};
            \node[]                     (topdotstext)  at ([xshift=0.2cm,yshift=0.5cm]topdots)  {\textcolor{gray!80!white}{$\overbrace{\phantom{this is a long, very long long text}}^{\lfloor\frac{|E|}{2}\rfloor\text{ edges}}$}};
            \node[node]                 (tu)        at ([xshift=1cm]topdots)  {};
            \node[]                     (tutext)   at ([yshift=-0.3cm]tu.center)  {$a'$};
            \node[node]                 (n3)        at ([xshift=1cm,yshift=-1cm]s)  {};
            \node[]                     (n3text)   at ([yshift=0.3cm]n3.center)  {$b$};
            \node[]                     (botdots)  at ([xshift=1cm]n3)  {\textcolor{gray!80!white}{\ldots}};
            \node[]                     (botdotstext)  at ([xshift=0.3cm,yshift=-0.1cm]botdots)
{\textcolor{gray!80!white}{$\underbrace{\phantom{this is a long, very long long text}}^{}$}};
            \node[]                     (botdotstexttext)  at ([yshift=-0.5cm]botdotstext)
{\textcolor{gray!80!white}{$^{\lfloor\frac{|E|}{2}\rfloor+1\text{ edges}}$}};
            \node[node, color=gray!80!white]                 (bu1)        at ([xshift=1cm]botdots)  {};
            \node[node]                 (bu2)        at ([xshift=1cm]bu1)  {};
            \node[]                     (bu2text)   at ([yshift=0.3cm]bu2.center)  {$b'$};
            \node[node]                 (d)        at ([xshift=0.5cm,yshift=1cm]bu2)  {};
            \node[]                     (dtext)   at ([yshift=0.3cm]d.center)  {$d$};
            
             \path  (s)    edge[normalArrow]    node[] {\tiny $\beta$}   (n2)
                            edge[normalArrow]   node[] {\tiny $\beta$}   (n3)
                    (n2)    edge[dottedArrow]   node[] {\tiny $\beta$}   (topdots)
                    (topdots.east)    edge[dottedArrow]   node[] {\tiny $\beta$}   (tu)
                    (tu)    edge[normalArrow]   node[] {\tiny $\beta$}   (d)
                    (n3)    edge[dottedArrow]   node[] {\tiny $\beta$}   (botdots)
                    (botdots.east)    edge[dottedArrow]   node[] {\tiny $\beta$}   (bu1)
                    (bu1)    edge[dottedArrow]   node[] {\tiny $\beta$}   (bu2)
                    (bu2)    edge[normalArrow]   node[] {\tiny $\beta$}   (d)
             ;
    \end{tikzpicture}
    \caption{Worst-case shortest paths topology}
    \label{fig:spworstcasegraph}
\end{figure}

\end{proof}

Now, let us consider what is the worst-case performance of a shortest widest (the shortest of the widest) paths algorithm for the SORR problem. Note that the worst-case scenario network topology differs from the shortest available paths algorithm as we try to provide the most difficult case for each individual strategy.

\begin{proposition}
The shortest widest paths algorithm has a competitive ratio of $\frac{1}{2}+\frac{1}{2+4\mu(|E|-1)}$.
\end{proposition}

\begin{proof}

(Direct proof.) When employing the shortest widest paths algorithm in a QKD network instance $N=(V,E,k)$, in the worst case it will affect as many other QKD requests as possible. Similarly to the shortest path analysis, in order for the widest path to impact the maximum number of other QKD requests, it should take all available keys in a path where many other assignments could have been made by an optimal algorithm. This can be easily achieved if QKD requests exist per each link used by the widest path assignment. 

Consider the QKD network instance depicted in Figure~\ref{fig:wpworstcasegraph}. Where for all links in exception of one the key capacity is constant, i.e., $k(e)=2\beta$, and a single link is of capacity $\beta$. In order to fail as many requests as possible, the idea is to make the shortest widest  path as long as possible, so that it affects a maximum number of other requests. This happens when the shortest widest path has a length of $|E|-1$ (The optimal algorithm would use the direct path which is not the widest.) This is evidently the case for networks with a \emph{ring}-like topology, as the alternative path must have a single link. 

Additionally, consider that for each edge in the shortest widest path there are $2\beta$ requests per link of a single bit, after $\frac{\beta}{\mu}$ requests of $\mu$ bits from $s$ to $d$. That is the sequence: {\small $\overbrace{(s,d,\mu)\ldots(s,d,\mu)}^{\frac{\beta}{\mu} \text{ times}}\overbrace{\overbrace{(s,b,1)\ldots(s,b,1)}^{2\beta \text{ times}}\ldots\overbrace{(b'd,1)\ldots(b',d,1)}^{2\beta \text{ times}}}^{2\beta(|E|-1) \text{ times}}$}. Note that this sequence would be completely served by an optimal algorithm, first it would use the direct (shortest and narrowest) path to serve the first $\frac{\beta}{\mu}$ requests from $s$ to $d$, and then, each individual request out of the $2\beta$ requests per link. For that reason, the competitive ratio of the shortest path algorithm is $\frac{\frac{\beta}{\mu}+\beta(|E|-1)}{\frac{\beta}{\mu}+2\beta(|E|-1)}$, simplifying the previous expression, we obtain that the shortest widest path algorithm has a competitive ratio of $\frac{1}{2+4\mu(|E|-1)}+\frac{1}{2}$. \qedhere

\begin{figure}[!htb]
    \centering
    \begin{tikzpicture}[node distance=1cm,>=stealth',bend angle=45,auto]
            \tikzstyle{node}=[circle,thick,draw=black!75, inner sep=0.075cm]
            \tikzstyle{normalArrow}=[thick,->]
            \tikzstyle{dashedArrow}=[thick,->,dashed]
            \tikzstyle{dottedArrow}=[thick,->,dotted, gray!80!white]

            \node[node]                 (s)       {};
            \node[]                     (stext)    at ([yshift=0.3cm]s.center)    {$s$};
            \node[node]                 (n3)        at ([xshift=1cm,yshift=-1cm]s)  {};
            \node[]                     (n3text)   at ([yshift=0.3cm]n3.center)  {$b$};
            \node[]                     (botdots)  at ([xshift=1cm]n3)  {\textcolor{gray!80!white}{\ldots}};
            \node[]                     (botdotstext)  at ([xshift=0.3cm,yshift=-0.1cm]botdots)
{\textcolor{gray!80!white}{$\underbrace{\phantom{this is a long, very long long text}}^{}$}};
            \node[]                     (botdotstexttext)  at ([yshift=-0.5cm]botdotstext)
{\textcolor{gray!80!white}{$^{|E|-1\text{ edges}}$}};
            \node[node, color=gray!80!white]                 (bu1)        at ([xshift=1cm]botdots)  {};
            \node[node]                 (bu2)        at ([xshift=1cm]bu1)  {};
            \node[]                     (bu2text)   at ([yshift=0.3cm]bu2.center)  {$b'$};
            \node[node]                 (d)        at ([xshift=0.5cm,yshift=1cm]bu2)  {};
            \node[]                     (dtext)   at ([yshift=0.3cm]d.center)  {$d$};
            
             \path  (s)     edge[normalArrow]   node[] {\tiny $2\beta$}   (n3)
                            edge[normalArrow]   node[] {\tiny $\beta$}   (d)
                    (n3)    edge[dottedArrow]   node[] {\tiny $2\beta$}   (botdots)
                    (botdots.east)    edge[dottedArrow]   node[] {\tiny $2\beta$}   (bu1)
                    (bu1)    edge[dottedArrow]   node[] {\tiny $2\beta$}   (bu2)
                    (bu2)    edge[normalArrow]   node[] {\tiny $2\beta$}   (d)
             ;
    \end{tikzpicture}
    \caption{Worst-case widest paths topology }
    \label{fig:wpworstcasegraph}
\end{figure}
\end{proof}

Our previous analyses seem to contradict the intuitive thought that by using the shortest paths we use more effectively the network resources, and thus, we could probably guarantee a better competitive ratio for the SORR problem. However, upon deep analysis, we showcase the contrary result. Furthermore, this helps shedding light into the important aspects of the SORR problem. For instance, using repeatedly the same links (as in shortest paths) may lead to quicker starvation of the resources. Additionally, by using the shortest widest paths, the network is left with overall more resources available for future requests, similar to a load-balancing strategy. 
Finally, it is important to note that the shortest available paths cannot be cached or parallelized, as the use of resources may alter the paths themselves. As a result, the shortest available paths scheme does not offer a clear computational complexity advantage over the widest-shortest paths approach.

\section{Experimental Evaluation}
This section aims at empirically confirming the theoretical results shown in the previous sections. Furthermore, our goal is not only to check the worst-case guarantees for online routing but, to assess the performance of both strategies in different settings; this includes on structured topologies and randomly generated ones, aiming at estimating real-world performance. 

\paragraph{Experimental setup}
Four different types of instances have been generated. Those are: the worst case shortest paths instances; the worst case widest path instances; hybrid random instances where the topology consists of a base star topology with a bus topology at the level-two nodes, and at least one full-mesh cluster at the third level, with random resource values, and random requests as described in \cite{pfar}; finally, fully random topologies, resources, and requests. For each of the different types of instances, 48 instances have been generated; ranging from 3 to 50 nodes. For each of these instances (4 categories, 192 in total), the widest paths algorithms, the shortest paths algorithm, and an idealized optimal implementation has been executed. The idealized optimal implementation works as an oracle, i.e., it has access to future requests, and based on it, it is capable of rejecting a request, differently from other algorithms. By executing this algorithms we obtain a service ratio, and by normalizing w.r.t. the optimal, we obtain the optimality ratio. An important note is that in all instances the request largely exceed the available capacity, to create resource starvation. 


\paragraph{Experimental results} Experimental results confirm our theoretical findings, and further give insights of the problem of interest. Figure~\ref{fig:wcwp} shows the results of both algorithms w.r.t. the worst case widest paths instances; as expected the widest paths show the $\frac{OPT}{2}$ worst-case guarantee. As can be seen this is a good instance for the shortest paths offering optimal performance. Also, as expected, the worst-case shortest paths instances (Figure~\ref{fig:wcsp}) show the close to zero performance for shortest paths, and a $\frac{OPT}{2}$ performance for the widest paths. As for the hybrid random (Figure~\ref{fig:hr}) and fully random (Figure~\ref{fig:fr}) instances we note that the $\frac{OPT}{2}$ guarantee for the widest paths algorithm cannot be given, as topologies with a single path between two requests must assign this path to the request whereas the optimal implementation is an idealized oracle, and given its future knowledge drops requests in order to maximize its service ratio. Moreover, a clear advantage can be seen for the widest-paths algorithm on random and hybrid random instances. We conjecture that shortest paths tends to overuse central links, which is undesirable when starvation occurs. We conclude that the widest path strategy has a clear performance advantage and is more stable across all experiments.
\begin{figure}[!htb]
    \centering
    \begin{tikzpicture}[scale=0.70]
    \begin{axis}[
        xlabel=Instance size (\# of nodes),
        ylabel=Optimality ratio,
        legend style={at={(0.35,0.01)}, anchor=south west},
        ymin=-0,
        ]
    
    \addplot[dotted,mark=*, mark size=0.8pt, black!40!white] plot coordinates {
      	(3,1.0)
		(4,1.0)
		(5,1.0)
		(6,1.0)
		(7,1.0)
		(8,1.0)
		(9,1.0)
		(10,1.0)
		(11,1.0)
		(12,1.0)
		(13,1.0)
		(14,1.0)
		(15,1.0)
		(16,1.0)
		(17,1.0)
		(18,1.0)
		(19,1.0)
		(20,1.0)
		(21,1.0)
		(22,1.0)
		(23,1.0)
		(24,1.0)
		(25,1.0)
		(26,1.0)
		(27,1.0)
		(28,1.0)
		(29,1.0)
		(30,1.0)
		(31,1.0)
		(32,1.0)
		(33,1.0)
		(34,1.0)
		(35,1.0)
		(36,1.0)
		(37,1.0)
		(38,1.0)
		(39,1.0)
		(40,1.0)
		(41,1.0)
		(42,1.0)
		(43,1.0)
		(44,1.0)
		(45,1.0)
		(46,1.0)
		(47,1.0)
		(48,1.0)
		(49,1.0)
		(50,1.0)
    };
    \addlegendentry{Shortest paths opt. ratio}
    \addplot[mark=*, mark size=0.8pt, black] plot coordinates {
      	(3,0.51220)
		(4,0.50820)
		(5,0.50617)
		(6,0.50495)
		(7,0.50413)
		(8,0.50355)
		(9,0.50311)
		(10,0.50276)
		(11,0.50249)
		(12,0.50226)
		(13,0.50207)
		(14,0.50192)
		(15,0.50178)
		(16,0.50166)
		(17,0.50156)
		(18,0.50147)
		(19,0.50139)
		(20,0.50131)
		(21,0.50125)
		(22,0.50119)
		(23,0.50113)
		(24,0.50108)
		(25,0.50104)
		(26,0.50100)
		(27,0.50096)
		(28,0.50092)
		(29,0.50089)
		(30,0.50086)
		(31,0.50083)
		(32,0.50081)
		(33,0.50078)
		(34,0.50076)
		(35,0.50073)
		(36,0.50071)
		(37,0.50069)
		(38,0.50067)
		(39,0.50066)
		(40,0.50064)
		(41,0.50062)
		(42,0.50061)
		(43,0.50059)
		(44,0.50058)
		(45,0.50057)
		(46,0.50055)
		(47,0.50054)
		(48,0.50053)
		(49,0.50052)
		(50,0.50051)
    };
    \addlegendentry{Widest paths opt. ratio}
    \end{axis}

\end{tikzpicture}
    \caption{Worst-case widest paths instances}
    \label{fig:wcwp}
\end{figure}

\begin{figure}[!htb]
    \centering
    \begin{tikzpicture}[scale=0.70]
    \begin{axis}[
        xlabel=Instance size (\# of nodes),
        ylabel=Optimality ratio,
        legend style={at={(0.35,0.99)}, anchor=north west},
        ]
    
    \addplot[dotted,mark=*, mark size=0.8pt, black!40!white] plot coordinates {
      	(3,0.091)
		(4,0.048)
		(5,0.048)
		(6,0.032)
		(7,0.032)
		(8,0.024)
		(9,0.024)
		(10,0.019)
		(11,0.019)
		(12,0.016)
		(13,0.016)
		(14,0.014)
		(15,0.014)
		(16,0.012)
		(17,0.012)
		(18,0.011)
		(19,0.011)
		(20,0.01)
		(21,0.01)
		(22,0.009)
		(23,0.009)
		(24,0.0082)
		(25,0.0082)
		(26,0.0076)
		(27,0.0076)
		(28,0.007)
		(29,0.007)
		(30,0.0066)
		(31,0.0066)
		(32,0.0062)
		(33,0.0062)
		(34,0.0058)
		(35,0.0058)
		(36,0.0055)
		(37,0.0055)
		(38,0.0052)
		(39,0.0052)
		(40,0.0049)
		(41,0.0049)
		(42,0.0047)
		(43,0.0047)
		(44,0.0045)
		(45,0.0045)
		(46,0.0043)
		(47,0.0043)
		(48,0.0041)
		(49,0.0041)
		(50,0.0040)
    };
    \addlegendentry{Shortest paths opt. ratio}
    \addplot[mark=*, mark size=0.8pt, black] plot coordinates {
		(3,1)
		(4,0.52381)
		(5,0.52381)
		(6,0.51613)
		(7,0.51613)
		(8,0.51220)
		(9,0.51220)
		(10,0.50980)
		(11,0.50980)
		(12,0.50820)
		(13,0.50820)
		(14,0.50704)
		(15,0.50704)
		(16,0.50617)
		(17,0.50617)
		(18,0.50549)
		(19,0.50549)
		(20,0.50495)
		(21,0.50495)
		(22,0.50450)
		(23,0.50450)
		(24,0.50413)
		(25,0.50413)
		(26,0.50382)
		(27,0.50382)
		(28,0.50355)
		(29,0.50355)
		(30,0.50331)
		(31,0.50331)
		(32,0.50311)
		(33,0.50311)
		(34,0.50292)
		(35,0.50292)
		(36,0.50276)
		(37,0.50276)
		(38,0.50262)
		(39,0.50262)
		(40,0.50249)
		(41,0.50249)
		(42,0.50237)
		(43,0.50237)
		(44,0.50226)
		(45,0.50226)
		(46,0.50216)
		(47,0.50216)
		(48,0.50207)
		(49,0.50207)
		(50,0.50199)
    };
    \addlegendentry{Widest paths opt. ratio}
    \end{axis}

\end{tikzpicture}
    \caption{Worst-case shortest paths instances}
    \label{fig:wcsp}
\end{figure}

\begin{figure}[!htb]
    \centering
    \begin{tikzpicture}[scale=0.70]
    \begin{axis}[
        xlabel=Instance size (\# of nodes),
        ylabel=Optimality ratio,
        legend style={at={(0.35,0.01)}, anchor=south west},
        ]
    
    \addplot[dotted,mark=*, mark size=0.8pt, black] plot coordinates {
      	(3,0.058)
		(4,0.122)
		(5,0.198)
		(6,0.225)
		(7,0.151)
		(8,0.217)
		(9,0.297)
		(10,0.278)
		(11,0.194)
		(12,0.269)
		(13,0.382)
		(14,0.281)
		(15,0.306)
		(16,0.309)
		(17,0.348)
		(18,0.341)
		(19,0.382)
		(20,0.434)
		(21,0.351)
		(22,0.455)
		(23,0.411)
		(24,0.448)
		(25,0.393)
		(26,0.376)
		(27,0.41)
		(28,0.381)
		(29,0.401)
		(30,0.508)
		(31,0.413)
		(32,0.497)
		(33,0.431)
		(34,0.463)
		(35,0.486)
		(36,0.515)
		(37,0.445)
		(38,0.507)
		(39,0.478)
		(40,0.533)
		(41,0.462)
		(42,0.431)
		(43,0.514)
		(44,0.503)
		(45,0.477)
		(46,0.477)
		(47,0.531)
		(48,0.541)
		(49,0.568)
		(50,0.573)
    };
    \addlegendentry{Shortest paths opt. ratio}
    \addplot[mark=*, mark size=0.8pt, black] plot coordinates {
      	(3,0.224)
		(4,0.303)
		(5,0.429)
		(6,0.493)
		(7,0.616)
		(8,0.512)
		(9,0.616)
		(10,0.626)
		(11,0.567)
		(12,0.621)
		(13,0.610)
		(14,0.566)
		(15,0.693)
		(16,0.660)
		(17,0.736)
		(18,0.642)
		(19,0.636)
		(20,0.700)
		(21,0.689)
		(22,0.734)
		(23,0.659)
		(24,0.732)
		(25,0.662)
		(26,0.755)
		(27,0.726829268292683)
		(28,0.703)
		(29,0.791)
		(30,0.772)
		(31,0.838)
		(32,0.742)
		(33,0.747)
		(34,0.730)
		(35,0.805)
		(36,0.757)
		(37,0.742)
		(38,0.761)
		(39,0.795)
		(40,0.702)
		(41,0.762)
		(42,0.735)
		(43,0.732)
		(44,0.748)
		(45,0.704)
		(46,0.755)
		(47,0.793)
		(48,0.758)
		(49,0.790)
		(50,0.794)
    };
    \addlegendentry{Widest paths opt. ratio}
    \end{axis}

\end{tikzpicture}
    \caption{Hybrid random instances}
    \label{fig:hr}
\end{figure}

\begin{figure}[!htb]
    \centering
    \begin{tikzpicture}[scale=0.70]
    \begin{axis}[
        xlabel=Instance size (\# of nodes),
        ylabel=Optimality ratio,
        legend style={at={(0.35,0.01)}, anchor=south west},
        ]
    
    \addplot[dotted,mark=*, mark size=0.8pt, black!40!white] plot coordinates {
		(3,0.26)
		(4,0.573)
		(5,0.53)
		(6,0.497)
		(7,0.579)
		(8,0.494)
		(9,0.487)
		(10,0.629)
		(11,0.669291338582677)
		(12,0.694)
		(13,0.562)
		(14,0.615)
		(15,0.718)
		(16,0.729)
		(17,0.770)
		(18,0.722)
		(19,0.843)
		(20,0.596)
		(21,0.608)
		(22,0.559)
		(23,0.596)
		(24,0.642)
		(25,0.622)
		(26,0.595)
		(27,0.61965811965812)
		(28,0.606)
		(29,0.621)
		(30,0.725714285714286)
		(31,0.565)
		(32,0.532558139534884)
		(33,0.497)
		(34,0.513)
		(35,0.46734693877551)
		(36,0.555)
		(37,0.669)
		(38,0.628)
		(39,0.621761658031088)
		(40,0.551053484602917)
		(41,0.597)
		(42,0.607)
		(43,0.5984375)
		(44,0.613)
		(45,0.584)
		(47,0.578)
		(48,0.470)
		(50,0.513595166163142)
    };
    \addlegendentry{Shortest paths opt. ratio}
    \addplot[mark=*, mark size=0.8pt, black] plot coordinates {
      	(3, 0.240000)
		(4, 0.573000)
		(5, 0.530000)
		(6, 0.497000)
		(7, 0.579000)
		(8, 0.494000)
		(9, 0.473000)
		(10, 0.629000)
		(11, 0.669291)
		(12, 0.620000)
		(13, 0.599000)
		(14, 0.717000)
		(15, 0.713000)
		(16, 0.707000)
		(17, 0.753138)
		(18, 0.722000)
		(19, 0.834000)
		(20, 0.735000)
		(21, 0.735000)
		(22, 0.726000)
		(23, 0.738000)
		(24, 0.742000)
		(25, 0.732000)
		(26, 0.771084)
		(27, 0.739000)
		(28, 0.817000)
		(29, 0.821000)
		(30, 0.890000)
		(31, 0.779000)
		(32, 0.772093)
		(33, 0.786000)
		(34, 0.813000)
		(35, 0.781633)
		(36, 0.821000)
		(37, 0.873000)
		(38, 0.882000)
		(39, 0.851000)
		(40, 0.853000)
		(41, 0.896000)
		(42, 0.910000)
		(43, 0.880000)
		(44, 0.891000)
		(45, 0.906000)
		(47, 0.897000)
		(48, 0.860000)
		(50, 0.896000)
    };
    \addlegendentry{Widest paths opt. ratio}
    \end{axis}

\end{tikzpicture}
    \caption{Fully random instances}
    \label{fig:fr}
\end{figure}

\section{Conclusion}
In this paper, we have presented a solution for the route planning, and online routing in quantum key distribution networks. For the route planing problem we propose a solution that is capable of deriving the appropriate routes for the demanded contracts. Furthermore, whenever this is unfeasible, we proposed a quadratic programming formulation, in order to obtain automated and \emph{fair} suggestions for the granted bandwidth for each of the contracts. With respect to the online routing problem, we have analyzed two different strategies, and have showcased both theoretically and empirically that the shortest path strategy performs poorly whereas the widest path strategy is able to guarantee worst-case performance of $\frac{OPT}{2}$ (where $OPT$ is the optimal algorithm performance). 

As for future work, we plan to complexify the route planning model in order to incorporate desirable notions, for instance, to guarantee a minimum granted bandwidth, and other constraints over the assigned paths. Furthermore, we are interested in looking for approximation algorithms or fixed parameter complexity algorithms that will guarantee a tractable running time as well as performance guarantees. Finally, we intend to look at the possibility of multi-path routing and the repercussions on such networks. As for the online routing problem, it is interesting to consider probability distributions of request and analyze the problem using average case analysis. Furthermore, it also interesting to collect data from  real-world implementations as these become available. Using these data, novel strategies (algorithms) based on evolutionary algorithms and large language models \cite{funsearch} can be derived.

\bibliographystyle{IEEEtran}
\bibliography{words/references}

@ARTICLE{qkdsdn,
  author={Aguado, Alejandro and Lopez, Victor and Lopez, Diego and Peev, Momtchil and Poppe, Andreas and Pastor, Antonio and Folgueira, Jesus and Martin, Vicente},
  journal={IEEE Communications Magazine}, 
  title={The Engineering of Software-Defined Quantum Key Distribution Networks}, 
  year={2019},
  volume={57},
  number={7},
  pages={20-26},
  doi={10.1109/MCOM.2019.1800763}}

@article{bb84++,
title = {Quantum cryptography: Public key distribution and coin tossing},
journal = {Theoretical Computer Science},
volume = {560},
pages = {7-11},
year = {2014},
note = {Theoretical Aspects of Quantum Cryptography – celebrating 30 years of BB84},
issn = {0304-3975},
doi = {https://doi.org/10.1016/j.tcs.2014.05.025},
url = {https://www.sciencedirect.com/science/article/pii/S0304397514004241},
author = {Charles H. Bennett and Gilles Brassard}
}

@misc{oqtavo,
  title={Towards a better approach for Quantum-Key-Distribution (QKD) Networks key management},
  author={Airbus, Secure Communications},
  howpublished = {\url{https://space-solutions.airbus.com/resources/news/secure-connectivity/quantum-key-distribution-networks-key-management/}},
  note = {Industrial communication, Last Accessed: 2025-08-13},
  year={2023}
}

@INPROCEEDINGS{mcf,
  author={Even, S. and Itai, A. and Shamir, A.},
  booktitle={16th Annual Symposium on Foundations of Computer Science (sfcs 1975)}, 
  title={On the complexity of time table and multi-commodity flow problems}, 
  year={1975},
  volume={},
  number={},
  pages={184-193},
  doi={10.1109/SFCS.1975.21}}

@article{qlinkperf,
	author = {Pirandola, Stefano},
	doi = {10.1038/s42005-019-0147-3},
	id = {Pirandola2019},
	isbn = {2399-3650},
	journal = {Communications Physics},
	number = {1},
	pages = {51},
	title = {End-to-end capacities of a quantum communication network},
	volume = {2},
	year = {2019}}

@INPROCEEDINGS{qkdssrouting,
  author={Tanizawa, Yoshimichi and Takahashi, Ririka and Dixon, Alexander R.},
  booktitle={2016 Eighth International Conference on Ubiquitous and Future Networks (ICUFN)}, 
  title={A routing method designed for a Quantum Key Distribution network}, 
  year={2016},
  volume={},
  number={},
  pages={208-214},
  doi={10.1109/ICUFN.2016.7537018}}

@INPROCEEDINGS{qkdssrouting2,
  author={Amer, Omar and Krawec, Walter O. and Wang, Bing},
  booktitle={2020 IEEE International Conference on Quantum Computing and Engineering (QCE)}, 
  title={Efficient Routing for Quantum Key Distribution Networks}, 
  year={2020},
  volume={},
  number={},
  pages={137-147},
  doi={10.1109/QCE49297.2020.00027}}

@article{qkdroutedemand,
	author = {Chen, Li-Quan and Zhao, Meng-Nan and Yu, Kun-Liang and Tu, Tian-Yang and Zhao, Yong-Li and Wang, Ying-Chao},
	date = {2021/09/14},
	doi = {10.1007/s11128-021-03246-2},
	id = {Chen2021},
	isbn = {1573-1332},
	journal = {Quantum Information Processing},
	number = {9},
	pages = {309},
	title = {ADA-QKDN: a new quantum key distribution network routing scheme based on application demand adaptation},
	url = {https://doi.org/10.1007/s11128-021-03246-2},
	volume = {20},
	year = {2021}}

@Article{qkdssrouting3,
AUTHOR = {Yao, Jiameng and Wang, Yaxing and Li, Qiong and Mao, Haokun and El-Latif, Ahmed A. Abd and Chen, Nan},
TITLE = {An Efficient Routing Protocol for Quantum Key Distribution Networks},
JOURNAL = {Entropy},
VOLUME = {24},
YEAR = {2022},
NUMBER = {7},
ARTICLE-NUMBER = {911},
URL = {https://www.mdpi.com/1099-4300/24/7/911},
PubMedID = {35885133},
ISSN = {1099-4300},
DOI = {10.3390/e24070911}
}

@Article{qkdprobarouting,
AUTHOR = {Bi, Lin and Miao, Minghui and Di, Xiaoqiang},
TITLE = {A Dynamic-Routing Algorithm Based on a Virtual Quantum Key Distribution Network},
JOURNAL = {Applied Sciences},
VOLUME = {13},
YEAR = {2023},
NUMBER = {15},
ARTICLE-NUMBER = {8690},
URL = {https://www.mdpi.com/2076-3417/13/15/8690},
ISSN = {2076-3417},
DOI = {10.3390/app13158690}
}

@INPROCEEDINGS{pfar,
  author={López, Jorge and Labonne, Maxime and Poletti, Claude and Belabed, Dallal},
  booktitle={2020 IEEE 19th International Symposium on Network Computing and Applications (NCA)}, 
  title={Priority Flow Admission and Routing in SDN: Exact and Heuristic Approaches}, 
  year={2020},
  volume={},
  number={},
  pages={1-10},
  doi={10.1109/NCA51143.2020.9306725}}

@article{funsearch,
  title={Mathematical discoveries from program search with large language models},
  author={Romera-Paredes, Bernardino and Barekatain, Mohammadamin and Novikov, Alexander and Balog, Matej and Kumar, M Pawan and Dupont, Emilien and Ruiz, Francisco JR and Ellenberg, Jordan S and Wang, Pengming and Fawzi, Omar and others},
  journal={Nature},
  volume={625},
  number={7995},
  pages={468--475},
  year={2024},
  publisher={Nature Publishing Group UK London}
}

\end{document}